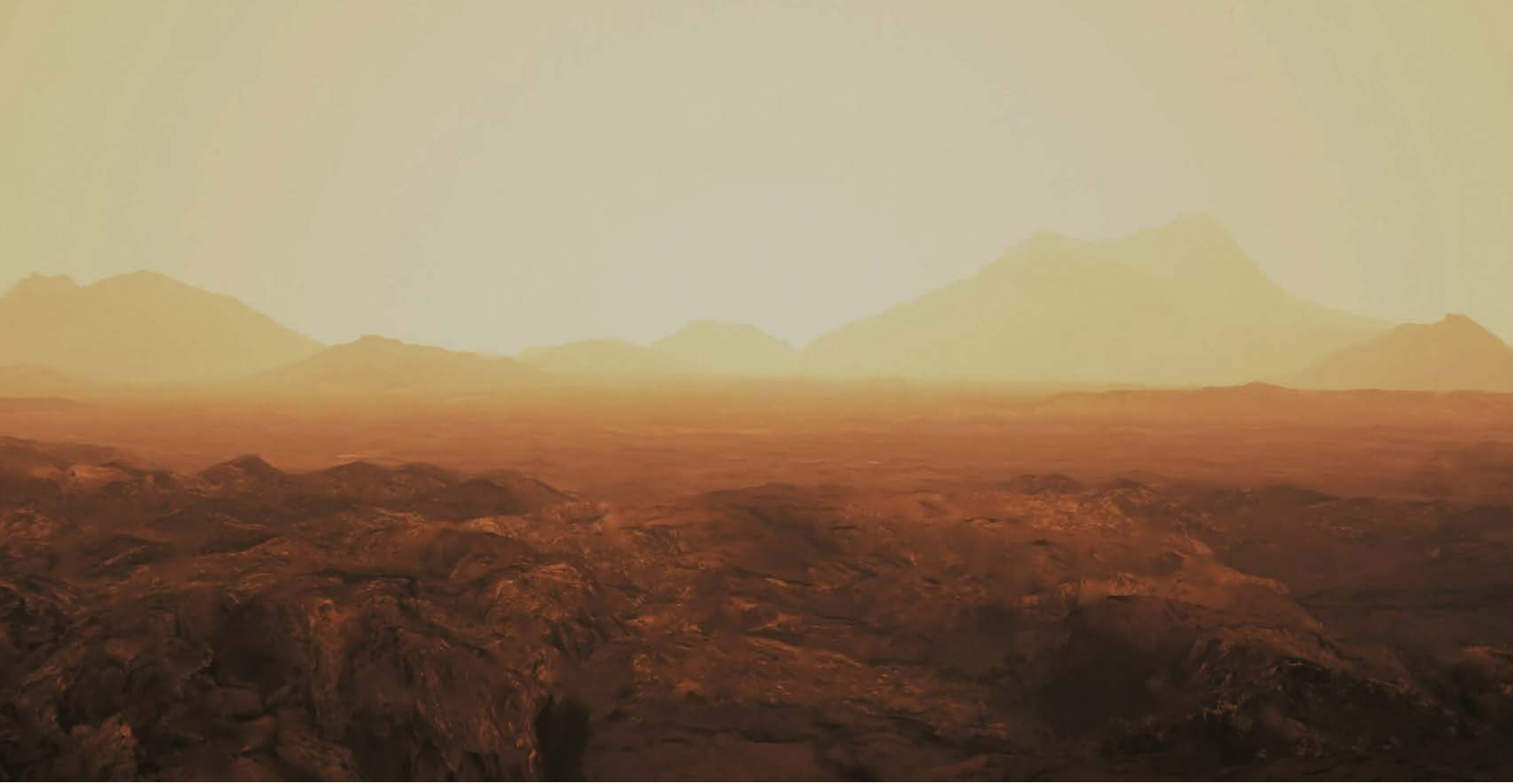

# Deep Atmosphere of Venus Probe as a Mission Priority for the Upcoming Decade

**AUTHOR:** James B. Garvin
NASA's Goddard Space Flight Center
Email: james.b.garvin@nasa.gov
Phone (cell): 301-646-4369

**CO-AUTHORS:**
Giada N. Arney (NASA GSFC)
Sushil Atreya (Univ. Michigan)
Stephanie Getty (NASA GSFC)
Martha Gilmore (Wesleyan)
David Grinspoon (PSI)
Natasha Johnson (NASA GSFC)
Stephen Kane (UC Riverside)
Walter Kiefer (LPI)
Ralph Lorenz (JHU/APL)

**CO-SIGNEES:**
Michael Amato (NASA GSFC)
Bruce Campbell (NASM, Smithsonian)
Dave Crisp (NASA JPL)
Scott Guzewich (NASA GSFC)
Sarah Horst (JHU/APL)
Noam Izenberg (JHU/APL)
Erika Kohler (NASA GSFC)
Paul Mahaffy (NASA GSFC)
Charles Malespin (NASA GSFC)
Alex Pavlov (NASA GSFC)
Mike Ravine (MSSS)
Jacob Richardson (NASA GSFC)
Melissa Trainer (NASA GSFC)
Chris Webster (NASA JPL)
Mikhail Zolotov (ASU)

*Deep Atmosphere of Venus Probe as a Mission Priority for the Upcoming Decade*

**EXECUTIVE SUMMARY**
*The deep atmosphere of Venus is largely unexplored and yet may harbor clues to the evolutionary pathways for a major silicate planet with implications across the solar system and beyond. In situ data is needed to resolve significant open questions related to the evolution and present-state of Venus, including questions of Venus' possibly early habitability and current volcanic outgassing. Deep atmosphere "probe-based" in situ missions carrying analytical suites of instruments are now implementable in the upcoming decade (before 2030), and will both reveal answers to fundamental questions on Venus and help connect Venus to exoplanet analogs to be observed in the JWST era of astrophysics.*

## INTRODUCTION AND BACKGROUND
### The Challenge of Measuring Venus' Massive Atmosphere

Previous Venus exploration has led to significant advancements in our understanding of the geodynamics and bulk atmospheric composition of the planet [Grinspoon & Bullock, 2007; Kane *et al.*, 2019; Way & Del Genio, 2020; Lammer *et al.*, 2020], even as profound questions remain such as those concerning atmospheric chemical stratification, possible signatures of present-day geologic and chemical activity, as well as Venus evolution. Compositional constraints from orbital near-IR night-side imaging have further produced new perspectives on the possible existence of "evolved" high-silica lithologies on Venus at scales of ~100 km [Hashimoto *et al.*, 2008; Gilmore *et al.*, 2015; Gilmore *et al.*, 2017; Weller & Kiefer, 2020], and high priority ancient terrains are ready to be interrogated. Plans for next-generation radar and night-side near IR emission spectrometers for mapping the surface at scales from tens of meters (SAR) to ~70 km (NIR) call for missions in the 2030s such as ESA's *EnVision* that will determine compositional patterns at regional to global scale for advancing models of Venus' crustal evolution [Ghail *et al.*, 2018].

Absence of compositional and dynamical information for the deepest atmosphere including noble gas, water, and major atmospheric isotopes and the basic T-P profile places limits on present understanding of Venus, such as outlined in Weller and Kiefer (2020), with connections to the role of past oceans [Donahue *et al.*, 1982; Kasting 1988; Way *et al.*, 2016; Way & Del Genio, 2020]. This fundamental knowledge gap can be effectively treated in the next decade via a new class of *in situ* mission, embracing key aspects of the VISE concept described in the 2011 NAS Decadal Survey [V&V, 2011] and recent VEXAG documents [VEXAG Goals, 2020]. Thanks to advancements in compact analytical instrumentation, high sensitivity descent imaging systems, and FPGA-enhanced probe flight systems, deep atmosphere "probes" are ready now to answer the top priority next questions posed by Venus in ways not possible for the past ~35 years. The case for such a "deep atmosphere probe with analytical chemistry" has been articulated since the 1983 Solar System Exploration Committee analysis [Morrison & Hinners, 1983] as the mission to follow the radar mapper *Magellan* and in all intervening planetary decadal surveys [NFSS, 2002; V&V, 2011]. Without definitive compositional measurements of Venus atmosphere down to its presently uninhabitable surface, advancements in models of its thermal and climate evolution will be impossible, thereby limiting the impact of our sister planet on our knowledge of solar system evolution (**Figure 1**). This mission White Paper describes the case for this class of mission concept (DEep Atmosphere Probe: DEAP) to resolve these knowledge gaps for the upcoming decade.

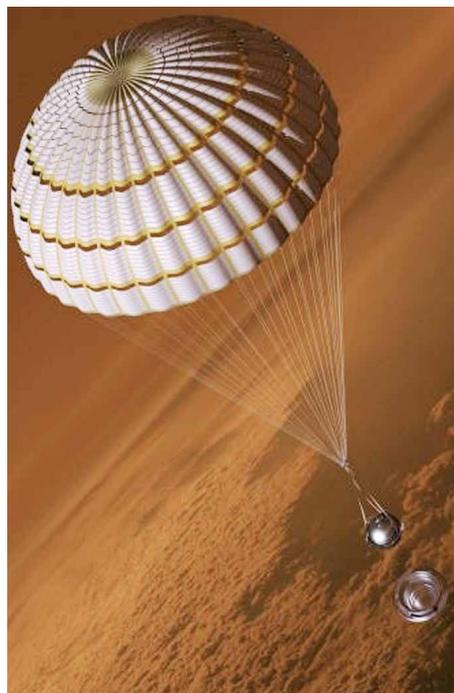

**Figure 1:** Conceptual Venus deep atmosphere probe mission for the 2020's decade. Such a mission could employ parachutes within the cloud deck (50-70 km) to enable time for gas ingest and processing and then freely fall to the surface at 10-15 m/s as it images the surface in the Near Infrared (NIR) windows to permit compositional mapping while profiling trace gases down to the surface (in their environmental context).

**Composition of the Venus Atmosphere: the Essential Next Step in Venus Exploration**

The Venus atmosphere holds clues to its origin, evolution, and dynamics and how it reflects the history of putative past oceans and active volcanism [Baines *et al.*, 2013; Bougher *et al.*, 1989; Treiman 2007; Garvin *et al.*, 2020]. The late-1970's measurements from Pioneer Venus (PVLP) were incomplete and did not offer the precision to measure the noble gases, especially Xenon and Helium [Lammer *et al.*, 2020], leaving ambiguities in our understanding of the planet. The single mid-atmosphere D/H value (~150) was suggestive of a large water inventory that was lost [Donahue *et al.*, 1982], but did not survey the variability in this key value from the top of the atmosphere to the near surface. No complete inventory of diagnostic trace gases was accomplished, especially for the deep atmosphere from ~16 km to the surface, where most (66%) of the atmosphere resides [Bougher *et al.*, 1989]. The lapse rate (temperature as a function of altitude) is insufficiently constrained and represents a key variable for current models of the deep atmosphere, where dominant $CO_2$ is super-critical [Lebonnois & Schubert, 2017]. No systematic compositional cross-section as a function of altitude from the mid-atmosphere clouds to the surface has ever been achieved. Without definitive compositional measurements of the bulk and lower-most Venus atmosphere, essential boundary conditions for evolutionary models that seek to explain Venus as a "system" cannot be developed [Kane *et al.*, 2019, **Figure 2**]. The composition of the near-surface atmosphere is needed to constrain the chemical alteration of surface materials and exchange of volatiles in the coupled atmosphere-surface rocks system [Zolotov, 2018, 2019]. Venus stands out as the least well-measured large atmosphere in the solar system (Lammer and others 2020), further limiting what our nearest neighbor planet can tell us about habitability of Earth-like planets and the broader workings of our solar system and planetary systems beyond [Kane *et al.*, 2019; NAS Exoplanets Strategy, 2018; Way *et al.*, 2016].

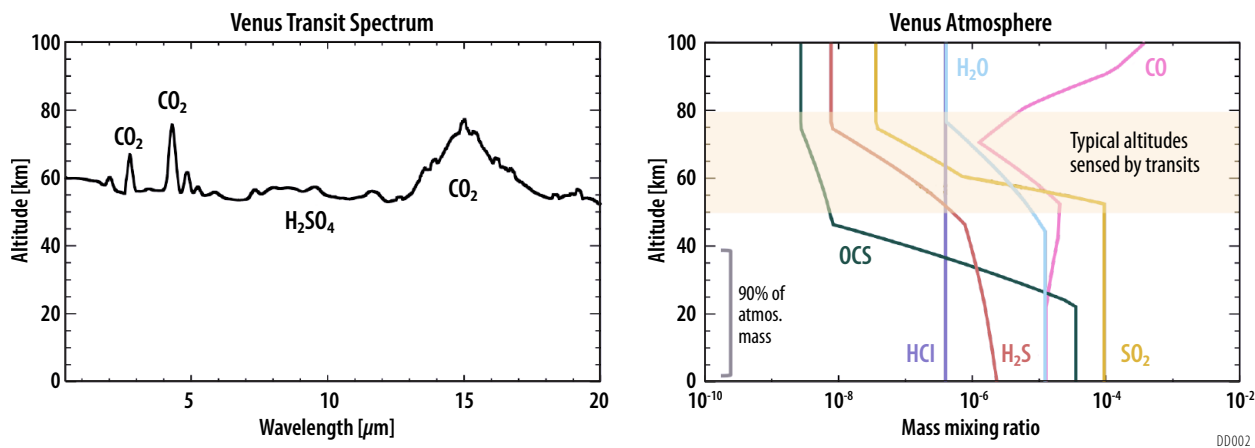

**Figure 2:** Greater understanding of Venus will provide higher fidelity simulations and data interpretation of what an exo-Venus might resemble to a future astrophysical observatory such as JWST or others planned for the 2030s versus the poorly-constrained Venus atmosphere (right), where most of the trace gas contents are uncertain, especially below ~45 km (*i.e.*, below the clouds) [Kane *et al.*, 2019]. A conceptual DEAP mission could survey details of the composition from 70 km to the surface to quantify what future transiting exoplanetary spectroscopy telescopes (JWST etc.) can evaluate beyond our solar system. Deep-atmosphere data is needed to constrain and validate models attempting to understand whole-atmosphere conditions of Venus-like exoplanets.

**The history of habitability at Venus?**

One of the most exciting emerging justifications for Venus cloud-deck compositional measurements are their critical role in assessing past or present habitability (and potential for biological activity). Cloud-deck microbial metabolism has become increasingly recognized as a significant venue for biology in a variety of Earth environments, including the stratosphere. Limaye and others (2018) summarized the case for scenario on present-day Venus, with specific indicator species such as phosphine ($PH_3$) as a detectable biosignature at Venus and in spectroscopy of exoplanets [Sousa-Silva *et al.*, 2020]. This scenario implies the *in situ* detectability of biogenic trace gases within the environmentally habitable cloud deck (~50-60 km altitude) today. Chemical signatures dating for Venus' oceanic period (or more recently) would be detectable with suitably sensitive analytical instrumentation of the type that have conducted related investigations on Mars as part of the Curiosity rover (as an example) for the past 8+ years [Trainer *et al.*, 2019]. Such instrumentation was largely non-existent 20 years ago but on the basis of investments that high sensitivity mass spectrometers (far exceeding the level of sensitivity enabled by remote sensing) on such missions as Curiosity (SAM) and Cassini/Huygens (INMS and GCMS), bringing such sensors to the "samples" throughout the Venus atmosphere (**Figure 1**) is now possible.

**GOALS AND RELEVANCE**

We define the comprehensive survey of the definitive composition, dynamics, and environment of the Venus atmosphere (~70 km to surface) as a primary science goal for Venus exploration in the next decade [V&V, 2011; VEXAG Goals, 2020]. Trace gases within the deepest atmosphere (*i.e.,* ~16 km to the surface), D/H in water, as well as Xe and He are particular targets of interest due to their relevance to climate, history, putative biology, and to surface weathering regimes [Fegley *et al.*, 1997; Zolotov, 2019]. Importantly, gradients in particular species across altitude bands (and hence as a function of pressure and temperature) provide insight into processes that connect the surface to the deep atmosphere over time scales relevant to major transitions [Weller & Kiefer, 2020; Way & Del Genio, 2020]. Thus, trace gas concentrations should be investigated across the entire sub-cloud atmosphere to dramatically extend current abundance data and improve understanding of the active mechanisms (thermochemical and photochemical reactions among gases, volcanism, chemical weathering, lateral variations) across the planet.

The comprehensive measurement of the Venus atmosphere is the logical next step from the perspectives of astrobiology and climatology [*e.g.*, Kane *et al.*, 2019; NAS Exoplanets Strategy, 2018]. Inventorying the distribution of altitude-dependent trace gases, as has been achieved at Titan and for Mars, is identified as a direct approach to detect volcanic activity and search for clues to major environmental transitions at regional to planetary scales. In-depth understanding of Venus' trace gases can help to determine the roles of known and unknown processes (*e.g.*, volcanic, thermochemical (gas-gas, gas-solid), photochemical, biological?) in controlling the current atmospheric composition to better constrain and refine relevant models beyond the current state-of-the-art [**Table 1**]. The updated 2020 VEXAG Goals state that determining sources and sinks of atmospheric trace gases are an essential objective for Venus, and in association with the New Frontiers "VISE" mission definition [V&V, 2011].

Significantly, definitive measurements of the bulk and trace atmosphere at relevant, representative altitudes (< 60 km) [Peplowski *et al.*, 2020] can provide missing boundary conditions for evolutionary models, as emphasized by Kane and others (2019) and Way and Del Genio (2020). As the VEXAG goals explicitly state, our knowledge of trace gas sources, sinks, and abundance across the entire Venus atmosphere remains far too limited. Additional measurements from beneath the cloud-deck to the surface, including near infrared compositional and "topographic" imaging [Garvin *et al.*, 2018, 2020], can connect the unknown trace gas contents and gradients to the local-to-regional geology of the surface in key regions that are themselves diagnostic of global-scale evolution in space and time [Weller & Kiefer, 2020], including complex ridged terrains. Picking up where Pioneer Venus Large Probe (PVLP) and the Soviet Venera and Vega landers left off, a deep atmosphere probe capable of profiling the trace gases from the clouds to the surface while measuring key environmental parameters and establishing ground truth in comparison to remote spectroscopy is amply justified.

**SCIENCE OBJECTIVES**

We define three primary science goals/objectives and associated measurements for quantifying the atmosphere of Venus from an *in situ* point of view in **Table 1** for a deep atmosphere probe (DEAP).

*Objective 1:* **Origin and Diversity of Atmosphere-bearing Planets**
Venus' atmosphere is an unexplored reservoir for comparing evolutionary pathways for large-atmosphere planets, and for connecting results to ongoing and upcoming studies of exoplanets, including those accessible to the James Webb Space Telescope (JWST).

*Objective 2:* **Evolution of Planetary Atmospheres and Habitability**
Understanding the history of water and other volatiles (including those involving S) for Venus from multi-altitude measurements of D/H and noble gas isotopes has the potential for transforming models of Venus oceanic state in space and time.

*Objective 3:* **Atmosphere/Surface Composition for Climate Relevance**

Compositional constraints on local to regional surface geology will address the role of water in both formation (*e.g.*, the role of water in the petrology of felsic rocks), as well as erosional and sedimentary processes that may have operated as tectonic regimes migrated over time on Venus.

**Table 1:** Connecting the current Planetary Decadal Survey [V&V, 2011] science goals and objectives to what a deep atmosphere probe at Venus (DEAP*) could provide in the upcoming decade. Suggested instruments from studies including Glaze *et al.* (2017) and Garvin *et al.* (2020). Please see also VEXAG Goals (2020) and text for details.

| NAS Decadal Inner Planet Science Goals | NAS Decadal Inner Planet Fundamental Science Objectives | New Frontiers VISE Objectives | DEAP* | Key Measurements | Example Instruments |
|---|---|---|---|---|---|
| Understand the Origin and Diversity of Terrestrial Planets | Constrain Bulk Composition of the Terrestrial Planets to Understand Formation from the Solar Nebula and Evolution | Understand the Physics and Chemistry of Venus' Atmosphere, Abundances of its Trace Gases, Sulfur, Light Stable Isotopes, and Noble Gas Isotopes | Meets | Noble gases & isotopes | QMS |
| | Characterize Planetary Surfaces to Understand Modification by Geologic Processes | Understand the Physics and Chemistry of Venus' Crust, and its possible evolutionary pathways across time | Partially Meets | Composition (surfaces) | NIR Descent Camera |
| | | Understand Weathering Environment of the Crust of Venus in the Context of Dynamics of the Atmosphere and the Composition and Texture of Surface Materials | Meets | Composition (deep atmos. & surface) | QMS + TLS + NIR descent camera |
| Understand How the Evolution of Terrestrial Planets Enables and Limits the Origin and Evolution of Life | Understand the Composition and Distribution of Volatile Chemical Compounds | Understand the Properties of Venus' Atmosphere down to the Surface and Improve Our Understanding of Zonal Cloud-Level Winds | Partially Meets | Trace gases and environmental context (p, T) | QMS with accelerometer and p, T sensors |
| Understand the Processes that Control Climate on Earth-Like Planets | Determine How Solar Energy Drives Atmospheric Circulation, Cloud Formation, and Chemical Cycles that Define the Current Climate on Terrestrial Planets | Constrain the Coupling of Thermochemical, Photochemical, and Dynamical Processes in Venus's Atmosphere and Between Surface and Atmosphere to Understand Radiative Balance, Climate, Dynamics, and Chemical Cycles | Meets | Trace gas composition and isotope ratios in C, H, N, O, S | QMS + TLS |
| | Characterize Record of and Mechanisms for Climate Evolution on Venus, with Goal of Understanding Climate Change on Terrestrial Planets, Including Anthropogenic Forcings on Earth | Look for Planetary-Scale Evidence of Past Hydrological Cycles, Oceans, and Life and for Constraints on Evolution of the Atmosphere of Venus | Meets | Noble and trace gas composition, isotope ratios, D/H, surface composition | NIR descent camera + TLS + QMS |
| | Constrain Ancient Climates on Venus and Search for Clues into Early Terrestrial Planet Environments to Understand Initial Conditions and Long-Term Fate of Earth's Climate | Indirect VISE Objective (but part of VEXAG 2020 goals) for New Frontiers | Meets | D/H, Noble gas composition and isotope ratios, trace gas composition, surface composition | QMS, TLS, NIR descent camera |

**PRIORITY ATMOSPHERIC ENTRY TARGET SITE TYPES**

As atmosphere-surface interactions are relevant globally across the surface of Venus [Zolotov 2018], there is a large degree of flexibility in choosing any site for entry-descent and atmospheric transect science with descent imaging for composition. A wide variety of highlands regions would be compelling from Maxwell Montes to isolated tesserae such as Alpha Regio and Tellus Regio [Gilmore *et al.*, 2015]. Revisiting previously investigated regions (PV probes, Venera, Vega landers) could also be beneficial for the purpose of building upon preexisting results. Given lack of chemical, surface texture, and lithology data for elevated highlands known as tesserae, having at least one deep atmosphere probe with near-surface imaging and trace gas composition would be desirable [Glaze *et al.*, 2017].

**MISSION STRATEGIES**

The proven track record of successful noble and trace gas measurements at Titan (Huygens) combined with well understood techniques for comprehensive atmospheric characterization in Martian environments [Curiosity SAM: Trainer *et al.*, 2019] make this a highly achievable mission concept for development in the next decade. A successful mission prioritizing atmospheric composition from ~70 km to the surface will require: (1) analytical instrumentation to measure gases with high sensitivity and signal-to-noise, (2) robust infrastructure to ingest, isolate, and process gases for measurement, (3) probe flight systems necessary to enable nadir-looking descent imaging for composition and topographic terrain analysis, and (4) entry systems to ensure an encapsulated probe flight system with suitable parachutes to enter and descend through the atmosphere over a surface region of interest at scales of ~350 km × 100 km (typical landing error ellipse at Venus from previous missions). On the basis of recent mission concept proposals and investments by spacefaring agencies, such a deep atmosphere probe flight system (**Figure 1**) with instruments and necessary avionics and telecommunications systems is implementable in the upcoming decade.

*(1) Detection Instrumentation:*

High-sensitivity noble and trace gas measurements, as have been demonstrated for over 8 years on Mars (and during probe descent at Titan) are available with flight-proven deep space experience, including Quadrupole Mass Spectrometers (QMS), Tunable Laser Spectrometers (TLS), and other varieties of gas-phase and aerosol analytical systems. **Table 1** documents traceability to science goals. Science objectives require ~1 ppbv limits of detection, with high-precision isotopic analytical capabilities for key atmospheric species.

*(2) Probe Infrastructure:*

Given past flight experience (Cassini's Huygens probe, and PV Large Probe), a deep atmosphere probe mission would require multiple redundant inlets for gas ingest and processing, necessary optical viewports (*e.g.*, sapphire) for instruments such as nadir-pointing descent imaging systems, penetrations to permit atmospheric structure and environmental measurements (p, T, accelerations), and additional penetrations for S-band radio-frequency telecommunications systems for data relay to supporting spacecraft (for ultimate downlink to Earth). These components have been demonstrated in planetary atmospheres (Mars, Venus, Titan, and Jupiter) and represent a low-technical risk approach for Venus.

*(3) Special Probe Flight System Support Requirements*

Beyond the capability of environmentally isolating analytical instruments from the Venus atmospheric environment for up to ~1.5 hours of descent, high optical through-put windows for descent imagers and other possible sensors (spectrophotometers) require relatively large port diameter (tens of mm). Descent imaging for composition and topography has yet to be performed at Venus, but experience from DISR on the Titan Huygens probe [Soderblom *et al.*, 2007] have demonstrated the potential of this approach. A nadir-pointed descent imaging system with near IR bands that permit surface radiance to be measured from below the clouds to the surface will permit discrimination of broad compositional trends ranging from high-silica rocks (felsic) to surfaces coated with alteration products such as hematite [Filiberto *et al.*, 2020; Zolotov 2018], and weathered basalts. **Figure 3** illustrates this potential at Venus, providing compositional assessment at scales from 5 m (at altitudes below ~2 km) to 100 m at higher altitudes (25 km). Combining compositional discrimination of end-member lithologies with 3D perspectives by processing of multi-frame, overlapping descent imaging [Garvin *et al.*, 2018] will provide first-of-its-kind geological characterization tied to lower atmosphere trace gas chemistry which may be in disequilibrium [Zolotov, 2018; Lebonnois & Schubert, 2017], thus, bringing lander/rover scale observations to Venus without the requirement for safe landings as a precursor to New Frontiers-class lander missions being considered for the 2030s [Garvin *et al.*, 2020].

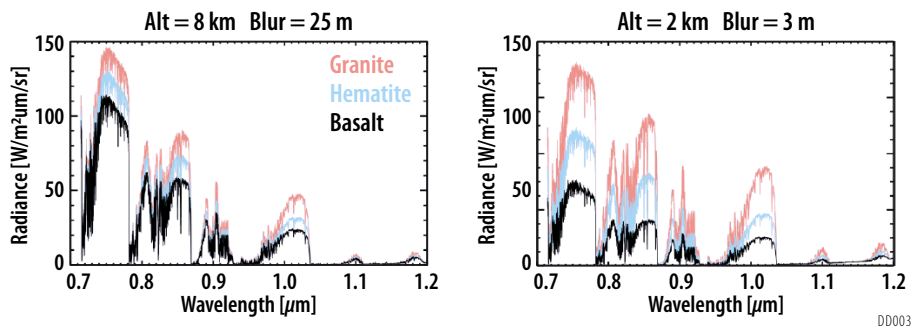

**Figure 3:** High-sensitivity descent imaging can discriminate between end-member rock types; 8 km (LEFT) and 2 km (RIGHT) altitudes shown, illustrating how readily felsic rocks can be distinguished even with the Rayleigh scattering and blur due to the massive Venus atmosphere. Mapping felsic rock units in the Venus highlands (tesserae) at scales from 100 m down to 5 m over areas as large as 25 x 25 km² is possible from a DEAP mission using descent camera technologies that have been demonstrated at Mars and Titan, and may be related to the role of water in rock formation and erosion [Gilmore *et al.*, 2015; Hashimoto *et al.*, 2008; Filiberto *et al.*, 2020].

**CONCLUSION**

Venus' atmosphere from the top of the cloud deck near 70 km to the surface presents a spectacular planetary laboratory that has remained largely unexplored. Ever since 1983 and the first prioritization of NASA planetary exploration missions [Morrison & Hinners, 1983], there has been a widely-recognized need for a deep atmosphere probe to Venus, which was echoed in the two most recent planetary Decadal Surveys. The case for such a mission in the decade of the 2020's is now more urgent as exoplanetary observations and models point to Venus-analog planets being commonplace beyond our solar system [NAS Exoplanets Strategy, 2018].